\documentclass[twocolumn,showpacs,preprintnumbers,amsmath,amssymb,prb]{revtex4}
\usepackage{graphicx}
\usepackage{dcolumn}
\usepackage{bm}
\usepackage{subfigure}

\begin{document}

\title{Antiferromagnetic topological insulator state in the correlated Bernevig-Hughes-Zhang model}

\author{S. Miyakoshi}
\author{Y. Ohta}
\affiliation{Department of Physics, Chiba University, Chiba 263-8522, Japan}

\date{15 March 2013}

\begin{abstract}
We study the effects of electron correlations on the topological phase 
transition in the Bernevig-Hughes-Zhang model using the variational cluster 
approach where the short-range spatial correlations are taken into account 
exactly.  We calculate the spin Chern number and local magnetic moment to 
show that the topologically nontrivial antiferromagnetic order exists and 
that the magnetic transition is of the second order.  We furthermore 
demonstrate that under the spin-quantized condition the topological phase 
transition is caused by the closing of the bulk band gap.  
\end{abstract}


\maketitle

\section{Introduction}
Presence of the topologically ordered phases\cite{wen,hatsugai} has received much 
attention in recent years from both theoretical and experimental 
sides.  The existence of gapless edge states and their characteristic 
electromagnetic responses detected in, e.g., the quantum Hall 
systems\cite{thouless,kohmoto} have been understood in terms of the topological orders 
and their unconventional insulator states distinguished from the 
simple insulator states by their nontrivial topological structure 
have been broadly referred to as the topological insulators (TIs).\cite{hasan,qi}  
Experimentally, the $\mathbb{Z}_{2}$ topological band insulators 
(TBIs),\cite{bernevig1,kane1,kane2,bernevig2,murakami,fu1,konig,hsieh,xia,chen} 
which are caused by the spin-orbit coupling without 
external magnetic field, have been realized in two- and 
three-dimensional systems such as HgTe/CdTe quantum wells\cite{bernevig2,konig} and 
bismuth based compounds.\cite{hsieh,xia,chen}  Their gapless edge states are protected by the 
time-reversal symmetry and their topology is reflected in the parity 
of the number of the edge states.\cite{schnyder,ryu,kitaev}  The relation between the 
crystallographic symmetry of the system and the presence of 
topologically ordered states in such systems has thus been studied 
intensively in recent years.\cite{fu2,mong,guo}  

The studies of the TBIs have been made within the noninteracting 
band theory on one hand.  However, the introduction of electron 
correlations to the TBIs has on the other hand attracted great 
attention from the viewpoint of the possible realization of exotic 
topological phases.\cite{shitade,pesin,gurarie,yamaji,hohenadler1,
yoshida1,yoshida2,yoshida3,tada,yu,wu,hohenadler2}  
In recent studies, a number of candidates 
for the exotic topologically ordered phases, where both the 
spin-orbit interaction and electron correlations play essential 
roles, have been proposed in transition-metal oxides, such as 
iridium oxides.\cite{shitade,pesin}  
For such systems, the properties of the TBIs, which reflect the 
effects of electron correlations and competition between the 
antiferromagnetic (AF) and topological orders, have been 
clarified in a wide variety of numerical methods, such as 
variational Monte Carlo (VMC) method,\cite{yamaji} 
quantum Monte Carlo (QMC) method,\cite{hohenadler1} 
dynamical mean-field theory (DMFT),\cite{yoshida1,yoshida2,yoshida3,tada} 
variational cluster approach (VCA),\cite{yu} and 
cluster dynamical mean-field theory (CDMFT).\cite{wu}  
In particular, in the DMFT study\cite{yoshida2} of the Bernevig-Hughes-Zhang 
(BHZ) model\cite{bernevig2} reproducing the TBI state, the coexistence of the TBI 
and axial antiferromagnetic Mott insulator (AFMI) state has been 
confirmed by including the on-site Hubbard interaction, the state 
of which is called the antiferromagnetic topological insulator 
(AFTI).  
In the QMC study\cite{hohenadler1} of the Kane-Mele model,\cite{kane1,kane2} 
another model reproducing the TBI state, it was found that the topological 
phase transition between the TBI and in-plane AFMI also appears but occurs 
without closing of the bulk band gap.  
The difference in the topological phase transitions between these 
two models has been understood as a consequence of the symmetry 
that conserves the $z$-component of the total-spin quantum numbers 
of the system, i.e., the spin-quantized condition.\cite{hohenadler2}  

However, because the spatial electron correlations cannot be 
treated sufficiently in the DMFT framework, the study of their 
effects on the AFTI states in the BHZ model still remains to 
be an important issue.  
In this paper, we will therefore apply the method of 
VCA\cite{potthoff1,potthoff2,senechal1,potthoff3,senechal2} to the 
BHZ model extended with the on-site Hubbard interaction, whereby 
we can treat the short-range spatial correlations exactly and 
calculate the single-particle Green function directly.  
We will thus evaluate the spin Chern number 
(SChN)\cite{yoshida1,yoshida2,yoshida3,volovik,fukui} and local 
magnetic moment of the AF order and show that the AF state actually 
has the topologically nontrivial structure, i.e., the AFTI state 
exists, under the spin-quantized condition.  
We will furthermore show that the magnetic transition between the 
TBI and AFTI is of the second order and will demonstrate that the 
topological phase transition between the AFTI and AFMI occurs 
with closing of the bulk band gap.  

This paper is organized as follows:  
In Sec.~II, we introduce the extended BHZ model with the 
on-site Hubbard interaction and briefly discuss the method 
of calculation.  In Sec.~III, we first calculate the 
properties of SChN and single-particle spectrum for the 
bulk and edge states in the nonmagnetic case and confirm 
that the result for SChN can determine the topological 
phase transition point in agreement with the single-particle 
spectra.  We then show the results for the AF ordered case 
and discuss the property of the topologically nontrivial 
AF order, including the influences on the single-particle 
spectra and local magnetic moments.  
A summary of the paper is given in Sec.~IV.  

\section{Model and method}

\begin{figure}[htbp]
\begin{center}
\includegraphics[width=0.7\columnwidth]{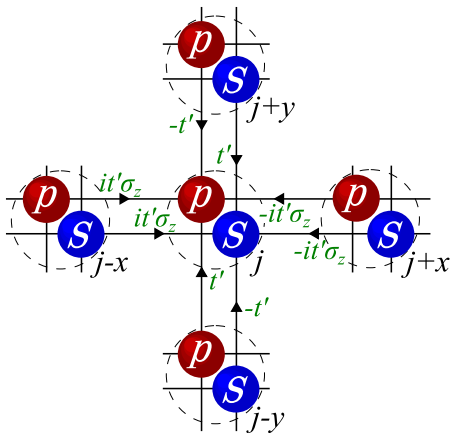}
\end{center}
\caption{(Color online) 
Schematic illustration of the spin-orbit coupling term in 
the BHZ model.  The circles labeled ``$s$'' and ``$p$'' 
correspond to the orbital $\alpha=1$ ($s$) and orbital 
$\alpha=2$ ($p$), respectively.  
$j$, $j\pm x$, and $j\pm y$ denote the lattice sites.  
}\label{BHZmodel}
\end{figure}

We study the BHZ model\cite{bernevig2} defined on the two-dimensional square 
lattice with an extension of including the on-site Hubbard 
interaction $U$.  The Hamiltonian reads
\begin{widetext}
\begin{subequations}
\begin{align}
{\cal H}&={\cal H}_{\rm BHZ}+{\cal H}_{\rm int}=\sum_{ij}\hat{c}^{\dagger}_{i}\hat{{\cal H}}_{ij}\hat{c}_{j}
+U\sum_{i\alpha}n_{i\alpha\uparrow}n_{i\alpha \downarrow} \\
\hat{{\cal H}}_{ij}&=
\begin{pmatrix}
\mathcal{H}_{ij}&0 \\
0&\mathcal{H}^{*}_{ij}
\end{pmatrix}
\\
\mathcal{H}_{ij}&=
\begin{pmatrix}
\epsilon_{1}\delta_{i,j}+t_{1}(\delta_{i,j \pm \hat{x}}+\delta_{i,j \pm \hat{y}}) 
& t'[i(\delta_{i,j+\hat{x}}-\delta_{i,j-\hat{x}})+\delta_{i,j+\hat{y}}-\delta_{i,j-\hat{y}}] \\
t'[i(\delta_{i,j+\hat{x}}-\delta_{i,j-\hat{x}})+\delta_{i,j-\hat{y}}-\delta_{i,j+\hat{y}}] 
& \epsilon_{2}\delta_{i,j}+t_{2}(\delta_{i,j \pm \hat{x}}+\delta_{i,j \pm \hat{y}}) 
\end{pmatrix}
\end{align}
\end{subequations}
\end{widetext}
where $\hat{c}_{i}=(c_{i1\uparrow},c_{i2\uparrow},c_{i1\downarrow},c_{i2\downarrow})^{\mathit{T}}$ and 
$n_{i\alpha\sigma}=c^{\dagger}_{i\alpha\sigma}c_{i\alpha\sigma}$.
This Hamiltonian contains two orbitals $\alpha=1,2$ and spin $\sigma=\uparrow,\downarrow$.  
The off-diagonal term in $\mathcal{H}_{ij}$, which corresponds to 
the spin-orbit coupling, is illustrated in Fig.~\ref{BHZmodel}.  
In our calculation, we consider the particle-hole symmetric case, 
assuming $-t_{1}=t_{2}=t$, $-\epsilon_{1}=\epsilon_{2}=t$, and $t'=0.8t$.

\begin{figure}[htbp]
\begin{center}
\includegraphics[width=0.6\columnwidth]{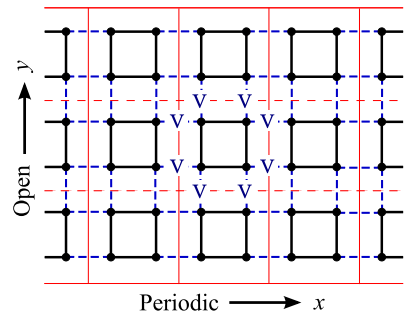}
\end{center}
\caption{(Color online) 
Sketch of the tiling used for calculating the single-particle 
spectrum for the edge.  The dotted line represents the 
inter-cluster hopping term $\mathbf{V}$.  We use the 4-site 
cluster for the bulk and edge states in the VCA calculations.  
To see the edge states, we arrange 10 clusters along the $y$ axis.  
}\label{edgeCPT}
\end{figure}

The VCA is a quantum cluster method based on the self-enegy 
functional theory (SFT)\cite{potthoff1,potthoff2} and has been applied to many 
interesting problems that include the competition between 
the symmetry-broken phases such as the superconducting 
phase, antiferromagnetic phase, etc.  
In the SFT framework, the problem of calculating the exact 
self-energy becomes a variational problem for the grand potential 
functional $\Omega[\mathbf{\Sigma}]$ defined as 
\begin{align}
\Omega[\mathbf{\Sigma}]=\mathrm{Tr\ ln}(\mathbf{G}^{-1}_{0}-\mathbf{\Sigma})^{-1}+F[\mathbf{\Sigma}]
\label{Omega}
\end{align}
where $F[\mathbf{\Sigma}]$ is the Legendre transfom of the 
Luttinger-Ward functional $\Phi[\mathbf{G}]$ such that the condition 
$\delta F[\mathbf{\Sigma}]=-T\mathbf{G} \cdot \delta \mathbf{\Sigma}$ 
is satisfied, and therefore, the condition 
$\delta \Omega[\mathbf{\Sigma}]=0$ corresponds to the Dyson equation 
$\mathbf{G}^{-1}=\mathbf{G}^{-1}_{0}-\mathbf{\Sigma}$.  
$\mathrm{Tr}$ in Eq.~\ref{Omega} represents both the sum over the 
Matsubara frequency and trace over the single-particle bases, 
such as momentum, spin, orbital, etc.  
In particular, the self-energy $\mathbf{\Sigma}$ can be obtained 
in VCA, where we use the reference system consisting of the small 
clusters with the single-particle parameters $\mathbf{t'}$, which are 
chosen so as to satisfy the stationary condition of the grand 
potential functional $\Omega[\mathbf{\Sigma'}(\mathbf{t'})]$.  
Furthermore, if we introduce the Weiss field as a variational parameter, 
we can treat the spontaneous symmetry breaking.  
In the VCA framework, the grand potential functional 
$\Omega[\mathbf{\Sigma'}(\mathbf{t'})]$ can be rewritten as 
\begin{align}
\Omega[\Sigma'(\mathbf{t'})]&=\Omega'(\mathbf{t'})-\mathrm{Tr\ ln}[\mathbf{G'}(\mathbf{t'})] \nonumber \\
&~~+\mathrm{Tr\ ln}[\mathbf{G}^{-1}_{0}-\mathbf{\Sigma'}(\mathbf{t'})]^{-1}
\end{align}
where $\Omega'(\mathbf{t'})-\mathrm{Tr\ ln}[\mathbf{G'}(\mathbf{t'})]$ 
corresponds to $F[\mathbf{\Sigma'}(\mathbf{t'})]$ in the reference system, 
and $\mathbf{G}_{0}$ is the free propagator of the original lattice system. 
Details of VCA can be found in Refs.~\onlinecite{potthoff3} and \onlinecite{senechal2}.  

In the evaluation of the SChN and single-particle spectrum, we use 
the cluster perturbation theory (CPT)\cite{senechal1} to calculate the lattice 
Green function $\mathbf{G}(\mathbf{t'})$, which is given by 
\begin{align}
\mathbf{G}(\mathbf{t'})&=[\mathbf{G}^{-1}_{0}-\mathbf{\Sigma'}(\mathbf{t'})]^{-1} \nonumber \\
&=[\mathbf{G}^{-1}_{0}-\mathbf{G'}^{-1}_{0}(\mathbf{t'})+\mathbf{G'}^{-1}_{0}(\mathbf{t'})
-\mathbf{\Sigma'}(\mathbf{t'})]^{-1} \nonumber \\
&=[-\mathbf{V}+\mathbf{G'}^{-1}(\mathbf{t'})]^{-1} \nonumber \\
&=\mathbf{G'}(\mathbf{t'})[\ \hat{1}-\mathbf{V}\mathbf{G'}(\mathbf{t'})]^{-1}
\end{align}
where $\mathbf{V}$ is the inter-cluster hopping treated 
perturbatively.  In the calculation of the bulk states, 
we use the supercluster where the clusters are arranged 
along the $x$ and $y$ axes and connected periodically 
along both directions.  For the edge states, on the other hand, 
we use the supercluster where the clusters are arranged 
along the $x$ and $y$ axes but connected periodically 
only along the $x$ axis (see Fig.~\ref{edgeCPT}).  

The SChN, which characterizes the topological property 
in the bulk states, may be written in the Green function 
formalism\cite{gurarie,yoshida1,yoshida2,volovik} as 
\begin{align}
N=\frac{\epsilon_{\mu \nu \rho}}{48\pi^{2}}\int{d^{3}k} \ 
\mathrm{Tr}[\hat{\sigma}_{z}\mathbf{G}(k)\partial_{\mu}\mathbf{G}^{-1}(k) \nonumber \\
\times \mathbf{G}(k)\partial_{\nu}\mathbf{G}^{-1}(k)\mathbf{G}(k)\partial_{\rho}\mathbf{G}^{-1}(k)]
\label{SChN}
\end{align}
where the notation $k=(i\omega,\mathbf{k})$ is used, 
$\partial_{\mu}$ denotes the partial differentiation 
with respect to $k^{\mu}$, and $\epsilon_{\mu\nu\rho}$ 
is the antisymmetric symbol.  This quantity is in proportion 
to the spin Hall conductivity and has the quantized value 
as long as the total spin $S^z_{\rm tot}$ of the system is 
conserved.\cite{ishikawa,haldane}  
In general, the $\mathbb{Z}_{2}$ topological insulator is 
characterized by $\mathbb{Z}_{2}$ invariant, which is 
well-defined as long as the time-reversal symmetry remains.  
The SChN on the other hand is equivalent to the parity 
of $\mathbb{Z}_{2}$ invariant on the condition of the 
spin conservation and is applicable to the axial AF case 
where the time-reversal symmetry is broken.  

Here, let us consider the SChN in the noninteracting BHZ model 
for simplicity.  In this case, SChN can be divided into the 
up- and down-spin parts as $N=N_{\uparrow}-N_{\downarrow}$, 
and the spin-polarized Hamiltonian $\hat{{\cal H}}_{\sigma}$ 
can be written in $\mathbf{k}$ space in the matrix form as
\begin{subequations}
\begin{align}
\hat{{\cal H}}_{\sigma}(\mathbf{k})&=\sum_{i}d_{i}(\mathbf{k})\hat{\tau}_{i} \label{Hsp0}\\
d_{1}(\mathbf{k})&=-2t'\mathrm{sin}\ k_{x}\times{\rm sign}(\sigma) \label{Hsp1}\\
d_{2}(\mathbf{k})&=-2t'\mathrm{sin}\ k_{y} \label{Hsp2}\\
d_{3}(\mathbf{k})&=-t-2t(\mathrm{cos}\ k_{x}+\mathrm{cos}\ k_{y}) \label{Hsp3}
\end{align}
\end{subequations}
where $\hat{\tau}_{i}$ represents the Pauli matrix for 
the orbital degrees of freedom.  Eq.~(\ref{SChN}) can then 
be rewritten in terms of the $\sigma$-spin contributions 
to the spin Chern number 
\begin{align}
N_{\sigma}=-\frac{1}{8\pi}\int{d^{2}\mathbf{k}}\ 
\hat{d}(\mathbf{k})\cdot \partial_{k_{x}}\hat{d}(\mathbf{k})
\times\partial_{k_{x}}\hat{d}(\mathbf{k})
\label{SChN2}
\end{align}
with $\hat{d}_{i}=d_{i}/|\hat{d}|$.  Using Eq.~(\ref{SChN2}), 
we obtain SChN as $N=1$ with $N_{\uparrow}=-N_{\downarrow}=1/2$, 
where the sign of SChN is determined from the sign of the spin-orbit 
interaction terms of Eqs.~(\ref{Hsp1}) and (\ref{Hsp2}) and reflects 
the direction of the current carried by the helical edge states.  

\begin{figure}[htbp]
\includegraphics[width=1.0\columnwidth,clip]{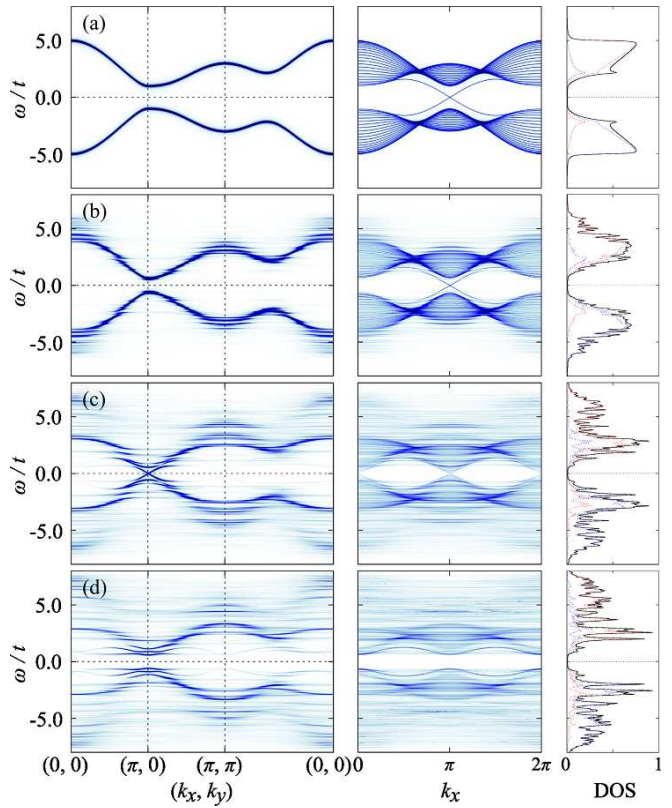}
\caption{(Color online) 
Single-particle spectra for the bulk (left panel) 
and edge state (middle panel) calculated by CPT.  
Right panel shows the density of states (DOS) for the bulk.  
The characteristic values of $U$ used are: 
(a) $U=0$ (TBI), 
(b) $U=2.0t$ (TBI), 
(c) $U=4.4t$ (topological phase transition point), and 
(d) $U=6.0t$ (NMI).  
} 
\label{nmiBand}
\end{figure}

\begin{figure}
\begin{center}
\includegraphics[width=0.8\columnwidth]{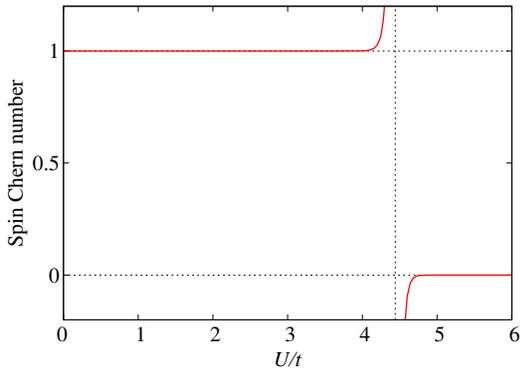}
\end{center}
\caption{(Color online) 
Calculated SChN for the nonmagnetic state.  
The vertical dotted line at $U=4.4t$ indicates the 
topological phase transition point between the 
TBI and NMI states.  
}\label{nmiSChN}
\end{figure}

\section{Results of calculation}

First, we discuss the effects of the on-site Hubbard 
interaction $U$ on the BHZ model in the nonmagnetic case.  
Figure \ref{nmiBand} shows the results for the 
single-particle spectra of the bulk and edge states. 
At $U<4.4t$, we find that the band gap in the bulk state 
(which we call the bulk gap) and the gapless edge 
state appear, which characterize the topologically 
nontrivial state, and thus we confirm that this region 
$U<4.4t$ belongs to the TBI state.  
Then, at $U=4.4t$, the bulk gap closes and simultaneously 
the gapless edge state vanishes (bulk-edge correspondence), 
indicating that the electron correlation $U$ derives the 
topological phase transition.  At $U>4.4t$, the bulk gap 
opens again but no gapless edge state appears, and thus the 
system belongs to the trivial nonmagnetic insulator (NMI) 
state.  
Interestingly, this phase transition occurs with closing 
the bulk gap and the single-particle spectrum for the 
bulk state acquires the linear dispersion.  This result 
has also been observed in the VCA study of the Kane-Mele 
model.  Therefore, in general, the closing of the bulk gap 
becomes a signal of the appearance of the topological 
phase transition between the TBI and NMI states.  

We note here that, although the single-particle spectra for 
the bulk and edge states are useful for detecting the 
topological phase transition as shown above, we cannot 
verify in this calculation whether the edge state actually 
brings the quantized spin current.  Also, it is not easy to 
take into account the possible inhomogeneity of the 
self-energies in the edge states in the VCA calculation.  
The SChN is free from such problems when we characterize 
the topology of the bulk states in correlated systems, 
which we will show in the following.  

The calculated result for SChN is shown in Fig.~\ref{nmiSChN}, 
where we find that, with increasing $U$, SChN discontinuously 
changes from 1 to 0 with a divergence at $U=4.4t$.  
This behavior corresponds to the bulk-gap closing at 
$U=4.4t$ shown in Fig.~\ref{nmiBand}.  In general, without 
any bulk-gap closing or anomaly of the self energy, the 
topological properties of the system do not change.\cite{gurarie,yoshida2}  
Therefore, this result also indicates that the topological 
phase transition between the TBI and NMI states occurs simultaneously 
with the bulk-gap closing.  In other words, this result 
supports the validity of our calculation of the single-particle 
spectra for the edge state.  
Note that our result for SChN does not show a step-function-like 
behavior from 1 to 0 but rather show a diverging behavior 
deviating from 1 or 0 in the vicinity of the transition point.  
This is due to a numerical error, which is however not serious 
for determining the topological phase transition point because 
the divergence of SChN signals the transition point accurately.  
This method of SChN can thus be used to determine the topological 
phase transition point between two topologically distinct states 
even when the system has the AF order, as we will discuss 
in the following.  

\begin{figure}
\begin{center}
\includegraphics[width=0.5\columnwidth]{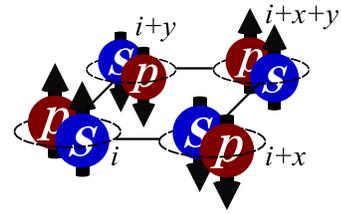}
\end{center}
\caption{(Color online) 
Schematic representation of the magnetic ordering in the 
BHZ model.  The spheres labeled ``$s$'' and ``$p$'' correspond 
to the orbital $\alpha=1$ (or $s$) and orbital $\alpha=2$ 
(or $p$), respectively, and the arrows indicate the spin 
directions.  
}\label{BHZaf}
\end{figure}

\begin{figure}
\begin{center}
\includegraphics[width=0.85\columnwidth]{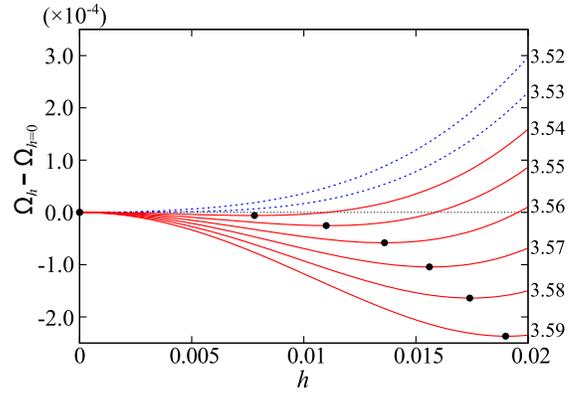}
\end{center}
\caption{(Color online) 
Calculated grand potential $\Omega_{\mathit{h}}-\Omega_{\mathit{h}=0}$ 
(per site) as a function of the Weiss field $h$ at several values of $U$.  
The value of $U$ used is shown on each curve.  The dots indicate the 
stationary points. 
}\label{SFT}
\end{figure}

\begin{figure}
\begin{center}
\includegraphics[width=0.9\columnwidth]{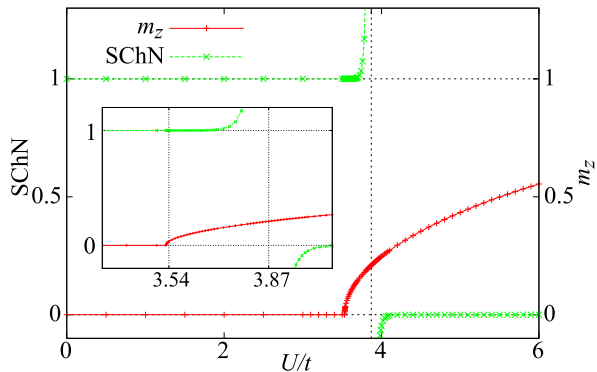}
\end{center}
\caption{(Color online) 
Calculated spin Chern number (SChN) and local magnetic momentum $m_{z}$.  
The vertical dotted line at $U$=3.87t indicates the topological phase 
transition point between the AFTI and AFMI states.  
The inset enlarges the region near the phase transition points.  
}\label{afSChN}
\end{figure}

\begin{figure}
\begin{center}
\includegraphics[width=1.0\columnwidth,clip]{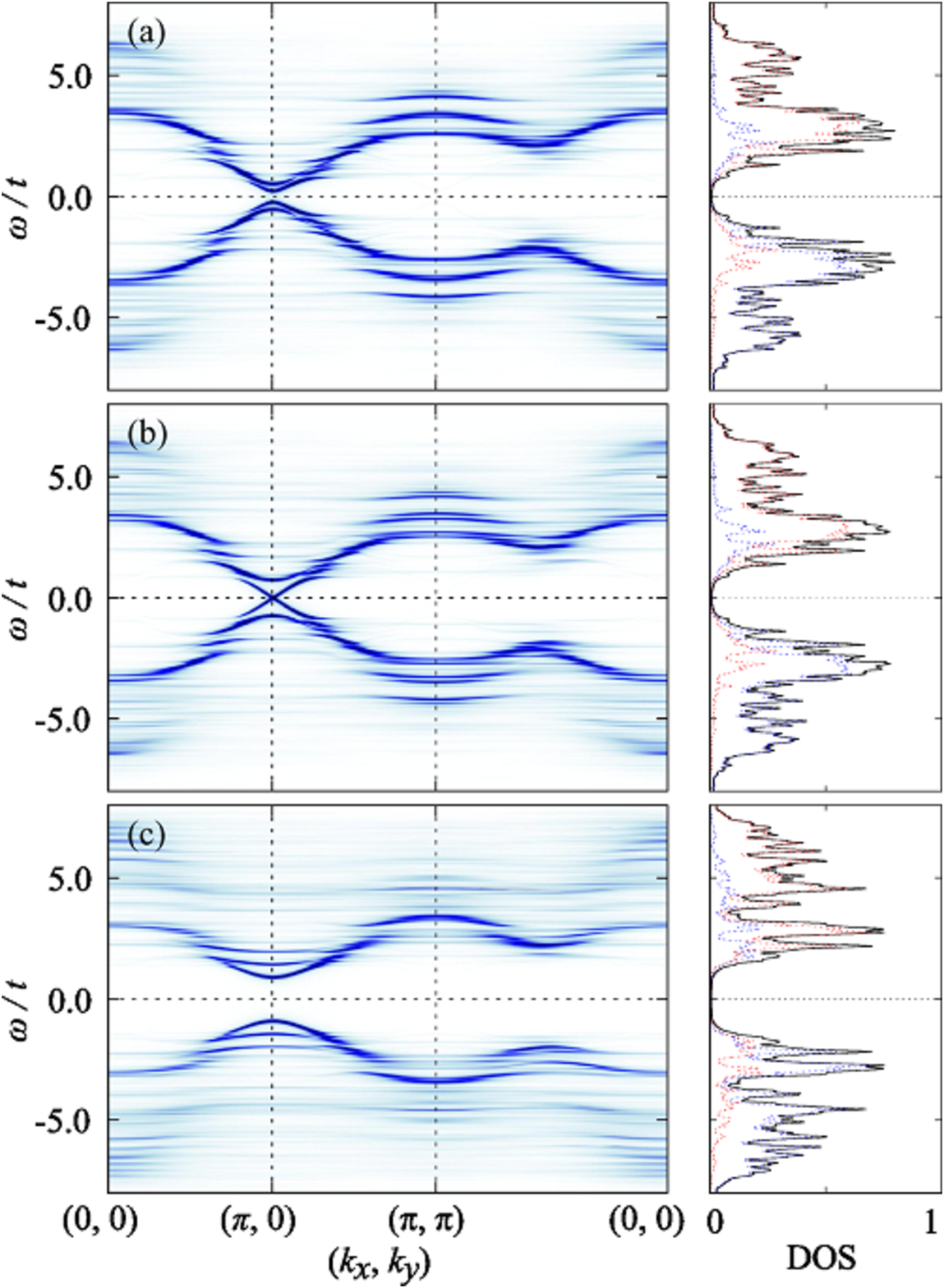}
\end{center}
\caption{(Color online) 
Calculated single-particle spectrum for the bulk (left panel) and 
corresponding DOS with the orbital decompositions (right panel) at 
(a) $U$=3.54t (magnetic transition point), 
(b) $U$=3.87t (topological phase transition point), and 
(c) $U$=6.00t (AFMI state).
}\label{afBand}
\end{figure}

Next, we discuss the magnetic instability in the BHZ model induced 
when we introduce the Hubbard interaction $U$.  
Let us first consider the case at large-$U$ limit, whereby 
we can search for candidates of the magnetic instabilities.  
Our model in this case may be rewritten by the following spin Hamiltonian: 
\begin{subequations}
\begin{align}
{\cal H}&=\sum_{\langle i,j\rangle \alpha \beta}\Bigl[J_{ij;\alpha\beta}
\mathbf{S}_{i\alpha}\cdot \mathbf{S}_{j\beta} \nonumber \\
& \ \ \ \ \ -2J^{xy}_{ij; \alpha \beta}
(S^{x}_{i\alpha}S^{x}_{j\beta}+S^{y}_{i\alpha}S^{y}_{j\beta})\Bigr]\\
J_{ij; \alpha \beta} &=\Bigl(\frac{4t^2}{U}\delta_{\alpha \beta}
+\frac{4t'^2}{U}(1-\delta_{\alpha \beta})\Bigr)(\delta_{i j\pm x}+\delta_{i j\pm y})\\
J^{xy}_{ij; \alpha \beta} &=\frac{4t'^2}{U}\delta_{i j\pm x}(1-\delta_{\alpha \beta})
\end{align}
\end{subequations}
with $\mathbf{S}_{i\alpha}=\frac{1}{2}\sum_{\tau,\tau'}c^{\dagger}_{i\alpha\tau}
[\hat{\boldsymbol{\sigma}}]^{\tau,\tau'}c_{i\alpha\tau'}$ 
and orbital indices $\alpha$ and $\beta$.  
This Hamiltonian has the in-plane ferromagnetic 
interaction $J^{xy}_{ij;\alpha\beta}$ coming from the 
spin-dependent anisotropic hopping term due to the spin-orbit 
coupling, and therefore, the spin space in 
this model does not have a full SU(2) symmetry but rather 
has an Ising-like axial anisotropy.  
In Fig.~\ref{BHZaf}, we show the spin configuration which 
is expected to occur in this Hamiltonian.  
This magnetic order satisfies the spin-quantized condition, 
and therefore, SChN is useful for determining the topological 
phase transition in our BHZ model under the magnetic order.  
No other magnetic orders, such as the in-plane orders, than 
that shown in Fig.~\ref{BHZaf}, have been detected, which 
is in contrast to the Kane-Mele model where the in-plane 
magnetic order occurs at the large $U$ limit.  
Note that the $\mathbb{Z}_{2}$ invariant is not meaningful 
due to the time-reversal symmetry breaking and cannot be used 
for determining the topological phase transition.  

Let us then assume the AF order shown in Fig.~\ref{BHZaf} 
and use the Weiss field of the form 
\begin{align}
{\cal H}_{\rm M}=h\sum_{i\alpha \tau \tau'}e^{i\mathbf{Q}\cdot \mathbf{r}_{i}}
c^{\dagger}_{i\alpha \tau}[\hat{\sigma}_{z}]^{\tau \tau'}c_{i\alpha \tau'}
\end{align}
with $\mathbf{Q}=(\pi, \pi)$ as the variational parameter 
in our VCA calculations.  
Figure \ref{SFT} shows the calculated result for the grand potential 
as the function of the Weiss field $h$ at several values of $U$.  
We find that there is a stationary point at $h \neq 0$ for 
$U>3.54t$ and thus the AF state appears for $U>3.54t$.  
We also calculate the local magnetic moment defined by 
\begin{align}
m_{Z}=\frac{1}{2N}\sum_{i\alpha \tau \tau'}
e^{i\mathbf{Q}\cdot\mathbf{r}_{i}}
\Big< c^{\dagger}_{i\alpha\tau}[\sigma^{z}]^{\tau\tau'}c_{i\alpha\tau'} \Big>
\end{align}
where $\langle\cdots\rangle$ represents the ground-state 
expectation value, $N$ is the number of the sites in the 
supercluster, and $\mathbf{Q}=(\pi, \pi)$.  

The calculated result for $m_Z$ is shown in Fig.~\ref{afSChN}, 
together with the result for SChN.  
We find that the magnetic transition occurs at $U=3.54t$, which 
is of the second order (contrary to the previous DMFT study\cite{yoshida2}), 
and that the SChN has the discontinuous change from 1 to 0 with the 
divergence at $U=3.87t$.  
We thus detect the region, i.e., $3.54t<U<3.87t$, where the topologically 
nontrivial state coexists with the AF order, which is nothing 
but the AFTI state.  

Figure \ref{afBand} shows the results for the single-particle 
spectra for the bulk state with the axial AF order at several 
values of $U/t$ $(=3.54,3.87,6.00)$.  
We find that the behaviors similar to the nonmagnetic case 
(see Fig.~\ref{nmiBand}) are observed also in the magnetically 
ordered case.  
We in particular confirm that the closing of the bulk gap is 
observed at $U=3.87t$ (see Fig.~\ref{afBand}(b)), i.e., at the 
topological phase transition point between the two topologically 
distinct phases, AFTI and AFMI.  
In the Kane-Mele model, it is known that the bulk gap does 
not close at the transition point between TBI and AFMI and 
that there is no coexistence region between the topological 
order and in-plane AF order.  Our results therefore indicate 
the importance of the the spin quantized condition, in relation 
to the symmetry in spin space, the coexistence region, and 
the bulk band gap.  

\section{Summary}

We have studied the effects of electron correlations on 
the topological phase transitions, focusing in particular on 
the presence of the antiferromagnetic topological insulator 
state.  We have used the Bernevig-Hughes-Zhang (BHZ) model 
with the on-site Hubbard interaction and have applied the 
variational cluster approach (VCA) whereby the short-range 
spatial correlations are taken into account exactly.  
We have calculated the spin Chern number (SChN) and local 
magnetic moment as well as the single-particle spectra for 
the bulk and edge states.  

In the nonmagnetic case, we found that the topological 
phase transition between the topological band insulator (TBI) 
and nonmagnetic insulator (NMI) occurs with the simultaneous 
vanishing of the bulk band gap and gapless edge state.  
This result may be a general conclusion equally valid 
for other topological phase transitions between TBI and 
NMI.  
We have also studied the magnetic instability of this model 
and have clarified the relation between the topological and 
antiferromagnetic (AF) orders.  We have calculated 
the spin Chern number and the local magnetic moment of the AF 
ordered state and confirmed that the magnetic transition, 
which was found to be of the second order, and topological 
phase transition do not occur simultaneously, but rather 
the topological order coexists with the AF order in the 
spin-quantized condition.  
From the single-particle gap for the bulk state, we have 
demonstrated that the topological phase transition between 
the AF topological insulator state and AF Mott insulator 
state occurs when the bulk gap closes.  

\begin{acknowledgments}
Enlightening discussions with S.~Ejima, H.~Fehske, Y.~Fuji, 
and K.~Seki are gratefully acknowledged.  
This work was supported in part by Kakenhi Grant 
No.~22540363 of Japan.  A part of computations was carried 
out at the Research Center for Computational Science, 
Okazaki Research Facilities, Japan.
\end{acknowledgments}

\end{document}